\documentclass[prd,aps,twocolumn,showpacs,preprintnumbers,amsmath,amssymb,
nofootinbib]{revtex4}
\usepackage{graphicx} 
\usepackage{graphics}
\usepackage{amssymb}
\usepackage{amsthm}

\def\beq{\begin{equation}}
\def\eeq{\end{equation}}
\def\bea{\begin{eqnarray}}
\def\eea{\end{eqnarray}}
\def\bec{\begin{center}}
\def\eec{\end{center}}
\def\half{\frac{1}{2}}
\def\nn{\nonumber}

\begin{document}

\date{\today}
\title{Particle Phenomenology on Noncommutative Spacetime}

\author{Anosh Joseph}\email[email: ]{ajoseph@phy.syr.edu}
\affiliation{Department of Physics, Syracuse University, Syracuse, NY 13244-1130 USA} 


\begin{abstract}
We introduce particle phenomenology on the noncommutative spacetime called the Groenewold-Moyal plane. The length scale of spacetime noncommutativity is constrained from the CPT violation measurements in $K^{0}-\bar{K}^{0}$ system and $g-2$ difference of $\mu^+ - \mu^-$. The $K^{0}-\bar{K}^{0}$ system provides an upper bound on the length scale of spacetime noncommutativity of the order of $10^{-32}$m, corresponding to a lower energy bound $E$ of the order of $E \gtrsim 10^{16}$GeV. The $g-2$ difference of $\mu^+ - \mu^-$ constrains the noncommutativity length scale to be of the order of $10^{-20}$m, corresponding to a lower energy bound $E$ of the order of $E \gtrsim 10^{3}$GeV. 
 
We also present the phenomenology of the electromagnetic interaction of electrons and nucleons at the tree level on the noncommutative spacetime. We show that the distributions of charge and magnetization of nucleons are affected by spacetime noncommutativity. The analytic properties of electromagnetic form factors are also changed and it may give rise to interesting experimental signals.
\end{abstract}
\maketitle

\section{Introduction}\label{intro}
Quantum field theories constructed on noncommutative spacetime provide a completely new perspective in building particle physics models beyond the Standard Model. (See \cite{AlvarezGaume:2003mb} for general properties of noncommutative field theories.) The noncommutative algebra of functions with the commutation relations between coordinate operators
\beq
\label{eq:comm}
[\hat{x}_{\mu}, \hat{x}_{\nu}] = i \theta_{\mu \nu},
\eeq
where $\theta_{\mu \nu}$ is a constant antisymmetric real matrix, can be used to model noncommutative spacetime. There are two major approaches to construct field theories on noncommutative spacetime. They are (i) the star-product formalism and (ii) the Seiberg-Witten map and enveloping algebra formalism. They both have a number of nice properties. We briefly discuss them below.

In the star-product formalism, to define a field theory on noncommutative spacetime, we replace the ordinary point-wise multiplication by a Moyal $\star$-product \cite{Weyl, Wigner, Moyal}
\beq
(f \star g)(x) = e^{\frac{i}{2}\theta_{\mu \nu}\partial_{x_{\mu}}\partial_{y_{\nu}}}f(x)g(y)\Big|_{x=y}.
\eeq

The replacement of point-wise products by star-products has several non-trivial consequences in field theories. The so-called UV/IR mixing \cite{Minwalla:1999px, Matusis:2000jf} is one such consequence. We can make the noncommutative field theory UV renormalizable by adding proper counter-terms. But we still have IR divergence problem. There have been a few proposals such as noncommutative hard resummation and (or) introducing a new way of regularization to resolve this problem \cite{VanRaamsdonk:2000rr, Griguolo:2001wg, Griguolo:2001ez}. The noncommutative gauge theories \cite{Martin:1999aq, Martin:2000bk, Bonora:2000ga}, the noncommutative version of real $\phi^4$ theory \cite{Minwalla:1999px, Aref'eva:1999sn, Micu:2000xj, Chepelev:1999tt, Chepelev:2000hm} as well as the complex $\phi^4$ theory \cite{Aref'eva:2000hq} and the noncommutative version of QED \cite{Hayakawa:1999yt, Riad:2000vy} have been shown to be one-loop renormalizable. In \cite{Chaichian:2001py} a noncommutative version of the Standard Model is constructed in the star-product approach. 

The presence of UV/IR mixing in noncommutative gauge theories in the star-product formalism makes the low energy physics of these theories to depend crucially on the details of ultraviolet completion. This can make the photons in the theory to have contributions from a trace-$U(1)$, causing vacuum birefringence (polarization dependent propagation speed). In \cite{Abel:2006wj} limits on the energy scale of noncommutativity are obtained from bounds on vacuum birefringence.

In the star-product approach with twisted Poincar\'e invariance and deformed statistics \cite{Chaichian:2004za, Balachandran:2005pn, Balachandran:2007kv, Balachandran:2007yf} it was shown that UV/IR mixing is present in a non-abelian gauge theory at one-loop level while it is absent in an abelian gauge theory to all orders of perturbative expansion \cite{Balachandran:2008gr}. On using this formalism it was shown that the power spectrum for cosmic microwave backgrond (CMB) becomes direction-dependent \cite{Akofor:2007fv} and a lower energy bound for noncommutativity parameter $\theta$ is obtained from the CMB data \cite{Akofor:2008gv}.   

An alternative method of noncommutative quantization was proposed in \cite{Madore:2000en} based on enveloping algebra valued fields and Seiberg-Witten maps. The introduction of noncommutativity in gauge theories limits the choice of gauge group to that of a matrix representation of a $U(N)$ gauge group. But we need a more general gauge group like $SU(N)$. The use of enveloping algebra valued fields seems to be the easiest way, but it can make the model meaningless since this would imply an infinite number of degrees of freedom \cite{Madore:2000en, Jurco:2000ja}. This problem can be solved by $\theta$-expanding the model using the so called Seiberg-Witten maps \cite{Seiberg:1999vs, Bichl:2001yf}. The Seiberg-Witten maps express noncommutative fields and parameters as local functions of the commutative fields and parameters. But problems still persist in the form of non-renormalizability of these theories. The $\theta$-expanded QED was shown to be power-counting non-renormalizable \cite{Wulkenhaar:2001sq}. Assuming that the noncommutative field theory under consideration is arising as the effective field theory of an unknown fundamental theory that is responsible for the noncommutativity of spacetime, the issue of renormalization can be tackled \cite{Calmet:2001na}. In \cite{Buric:2006wm} the gauge sector of the noncommutative standard model was shown to be one-loop renormalizable to first order in the expansion in $\theta$. (See also \cite{Grimstrup:2002af, Latas:2007eu, Martin:2007wv, Buric:2007ix}.) In \cite{Bichl:2001cq} it was shown that the photon self-energy is renormalizable to all orders in this approach. 

Noncommutative non-abelian gauge theories are constructed in \cite{Jurco:2001rq}. See \cite{Calmet:2001na, Melic:2005fm, Melic:2005am} for the constructions of the Standard Model on noncommutative spacetime using this approach.

Noncommutative gauge theories formulated using this approach may contain additional gauge anomalies which do not appear in ordinary commutative gauge theories. In \cite{Brandt:2003fx} it was shown that noncommutative gauge theories with arbitrary compact gauge group have the same one-loop anomalies as their commutative counterparts. (See also \cite{Martin:2002nr}.)

There has been much progress in applying Seiberg-Witten map based noncommutative field theories in the context of high energy physics phenomenology of noncommutative Standard Model \cite{Behr:2002wx, Duplancic:2003hg, Melic:2005hb, Melic:2005su, Alboteanu:2006hh, Buric:2006nr, Alboteanu:2007bp, Trampetic:2008bk, Conley:2008kn, Tamarit:2008vy}, noncommutative neutrino physics \cite{Schupp:2002up, Minkowski:2003jg, Horvat:2008uc}, astrophysics \cite{Haghighat:2009pv} and cosmology \cite{Horvat:2009cm, Ettefaghi:2009ai}.

In this paper we focus on the particle phenomenology of noncommutative field theories constructed using the star-product formalism along with twisted statistics \cite{Balachandran:2005pn, Balachandran:2007kv, Balachandran:2007yf}. In \cite{Chaichian:2004za} it was shown that noncommutative spacetime with the commutation relations given in Eq. (\ref{eq:comm}) can be interpreted in a Lorentz invariant way by invoking the concept of twisted Poincar\' e symmetry of the algebra of functions on a Minkowski spacetime. Twisting of the Poincar\'e symmetry leads to twisted statistics in noncommutative field theories \cite{Balachandran:2005pn, Balachandran:2006pi, Balachandran:2007kv, Balachandran:2007yf, Balachandran:2007vx, Akofor:2008ae, Balachandran:2008gr}.

Twisted noncommutative quantum field theories are shown to be Lorentz non-invariant \cite{Balachandran:2008gr, Akofor:2008ae, Balachandran:2007kv, Balachandran:2007yf, Balachandran:2007vx, Balachandran:2006pi, Balachandran:2005pn, Chaichian:2004yh, Chaichian:2004za}, CPT violating \cite{Akofor:2007hk} and non-local in nature \cite{Balachandran:2007yf, Akofor:2008gv, Soloviev:2007hp, Kobakhidze:2008cq}. The scattering matrix of these theories cannot be Lorentz invariant in general. They can depend upon the noncommutativity parameter $\theta^{\mu \nu}$ and the external four-momenta of the scattering particles. The incident- and outgoing state vectors are also modified by the spacetime noncommutativity. The frame dependence of $S$-matrix gives rise to many interesting features in such quantum field theories.  It gives rise to nontrivial effects such as corrections to the electric and magnetic properties of nucleons. In a general frame, the appearance of $\theta^{\mu \nu}$ in the $S$-operator can also break the discrete symmetries P and CPT \cite{Akofor:2007hk}.   

We organize the paper as follows: In section \ref{kaon-system}, we constrain the noncommutativity parameter from the CPT violation measurements in the neutral kaon ($K^{0}-\bar{K}^{0}$) system \cite{Gibbons1997}. In section \ref{g-minus-two}, we put further constraint on the noncommutativity length scale from the CPT violation measurements on the $g-2$ of $\mu^+ - \mu^-$ \cite{:2007yc, Bailey, Carey, Brown}. We follow a phenomenological approach without invoking the microscopic structure of the underlying theory in detail. In section \ref{form-factors}, we show how the electron-nucleon scattering process at the tree level is affected by spacetime noncommutativity. The vertex function and thus the electromagnetic form factors of the nucleon are modified. They depend on the total incident- and recoil four-momenta, and the noncommutativity parameter. Analytic properties of electromagnetic form factors are also changed, indicating the possibility of experimental signals. The paper concludes in section \ref{conclusions}. 
\section{The neutral kaon system in noncommutative framework}\label{kaon-system}
\begin{figure}
\begin{center}
\includegraphics{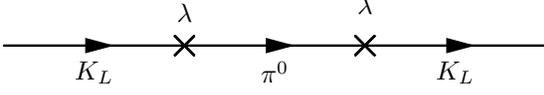}
\caption{A self-energy diagram for $K_{L}$ with $\pi^{0}$ pole-dominance. The coupling constant $\lambda$ has dependence on $\vec{\theta}^{0}$ and the total incident momentum $\vec{P}_{in}$.} \label{fig:mass-renorm}
\end{center}
\end{figure}

In the standard (commutative) case, if we start off with a pure $K^{0}$ beam created in a strong interaction process at time $t=0$, its intensity $I(K^{0})$ oscillates with a frequency $\Delta m \equiv m_{L} - m_{S}$, where $m_{L}$ ($m_{S}$) is the mass of the weak-interaction kaon eigenstate $K_{L}$ ($K_{S}$). When kaons propagate through space they are distinguished by their mode of decay and thus by the weak-interaction eigenstates \cite{Perkins2000}. For $K_S$, the lowest mass intermediate states are two-pion states and they are expected to be smaller than one-pion intermediate state that occurs for $K_L$ mass renormalization (see Fig. \ref{fig:mass-renorm}). We infer that $K_{L}$ should be affected the most by spacetime noncommutativity.

In an arbitrary scattering diagram involving quark-quark-gluon ($q-q-g$) vertices, the  space-space part of noncommutativity can be integrated out to zero from the $S$-operator \cite{Balachandran:2007yf}. The $S$-operator carries the time-space part of the noncommutativity through the twist factor \cite{Balachandran:2007yf, Drinfeld1990, Chaichian:2004yh, Chaichian:2004za} $\exp\Big(\frac{1}{2}\overleftarrow{\partial}_{0}\vec{\theta}^{0}.\vec{P}_{in}\Big)$, where $\overleftarrow{\partial}_{0}$ differentiates the appropriate time argument and $\vec{\theta}^{0} = (\theta^{01}, \theta^{02}, \theta^{03})$.

The self-energy diagram for $K_{L}$ with one-pion pole dominance should be affected by this twist factor. Such a self-energy diagram is depicted in Fig. \ref{fig:mass-renorm}. The coupling constant $\lambda$ has dependence on $\vec{\theta}^{0}$ and the total incident momentum $\vec{P}_{in}$ through the $q-q-g$ vertices appearing in the microscopic version of the theory. We will not discuss the details of the theory in the quark-gluon level as it is far too complicated.  

The non-local nature of the theory in time and thus the appearance of $\vec{\theta}^{0}$ in scattering processes as mentioned above indicates that we should twist the mass and decay width of $K_{L}$:
\bea
m_{L}^{\theta} &=& m_{L}\exp\Big(\frac{i}{2}m_{K^{0}}\vec{\theta}^{0}.\vec{P}_{in}\Big),\\
\gamma_{L}^{\theta} &=& \gamma_{L}\exp\Big(\frac{i}{2}m_{K^{0}}\vec{\theta}^{0}.\vec{P}_{in}\Big),
\eea
where $m_{L}$ and $\gamma_{L}$ are the mass and width of $K_{L}$ eigenstate in the commutative case and $m_{K^{0}}$ is the mass of strong interaction eigenstate of $K^{0}$. Notice that these expressions recover their respective commutative forms when $\vec{\theta}^{0}=0$, $\vec{P}_{in}=0$ or $\vec{\theta}^{0}.\vec{P}_{in}=0$. 

We can obtain an expression for the $K^{0}$ intensity at time $t$ from the amplitudes of the states $K_{S}$ and $K_{L}$:
\bea
A_{S}(t) &=& A_{0}\exp\Big[(-\frac{1}{2}\gamma_{S} + im_{S})t\Big], \\
A_{L}(t) &=& A_{0}\exp\Big[(-\frac{1}{2}\gamma^{\theta}_{L} + im^{\theta}_{L})t\Big],
\eea
where $A_0$ is the amplitude at time $t=0$.

If the $K^{0}$ beam is pure with unit intensity when $K^0$'s are created at $t=0$, we have the expression for intensity at time $t$,
\begin{widetext}
\bea
I_{K^{0}} &=& \frac{1}{2}[A_{S}(t) + A_{L}(t)][A^{*}_{S}(t) + A^{*}_{L}(t)]\nn \\
\label{eq:intensity}
&=& \frac{1}{4}\Big[\textrm{exp}(-\gamma_{S}t)+\textrm{exp}\Big(-\gamma_{L}\cos \alpha t+2 m_{L}\sin \alpha t\Big)+2\exp\Big(-\half\gamma_{S}t-\half\gamma_{L}\cos \alpha t + m_{L}\sin \alpha t\Big)\cos\Delta m t\Big],
\eea
\end{widetext}
where
\beq
\Delta m = \frac{\gamma_{L}}{2}\sin \alpha + m_{L}\cos \alpha - m_{S},
\eeq
and 
\beq
\alpha = \half m_{K^{0}}\vec{\theta}^{0}.\vec{P}_{in}.
\eeq

From Eq. (\ref{eq:intensity}), we can define a width difference
\beq
\Delta \gamma = \gamma_{S}-\gamma_{L}\cos \alpha + 2 m_{L}\sin \alpha .
\eeq
These expressions also recover their standard forms in the $\theta^{\mu \nu} \rightarrow 0$ limit. 

If we assume that the mass and width differences are arising purely due to spacetime noncommutativity, then we have $m_{L}-m_{S}=0$ and $\gamma_{S}-\gamma_{L}=0$ for the case $\theta^{\mu \nu}=0$. In that case, $m_{L}=m_{S}=m_{K^{0}}$ and $\gamma_{S} = \gamma_{L}= \gamma_{K^{0}}$. Then we have the noncommutative expressions to the lowest order in $\vec{\theta}^{0}$:
\bea
\label{eq:lowest-order1}
\Delta m &\simeq& \frac{\gamma_{K^{0}}}{2}~(\frac{m_{K^{0}}}{2}\vec{\theta}^{0}.\vec{P}_{in}),\\
\label{eq:lowest-order2}
\Delta \gamma &\simeq& 2 m_{K^{0}}~(\frac{m_{K^{0}}}{2}\vec{\theta}^{0}.\vec{P}_{in}).
\eea

In the standard phenomenological theory of kaons, the CPT violation complex parameter $\delta$ is defined as
\bea
\label{eq:delta-eqn}
\delta &=& \frac{\Lambda_{\bar{K}^{0}\bar{K}^{0}}-\Lambda_{K^{0}K^{0}}}{2(\lambda_{L}-\lambda_{S})}\nn \\
&=& \delta_{\|}\exp(i\phi_{SW}) + \delta_{\bot}\exp(i(\phi_{SW}+\frac{\pi}{2})),
\eea
where $\Lambda_{\bar{K}^{0}\bar{K}^{0}}$ and $\Lambda_{K^{0}K^{0}}$ are diagonal entries of the $2 \times 2$ matrix $\Lambda \equiv M -\frac{i}{2}\Gamma $, $\lambda_{L,S} = m_{L,S}-\frac{i}{2}\gamma_{L,S}$ are the eigenvalues of the matrix $\Lambda$, $\delta_{\|}$ and $\delta_{\bot}$ are respectively the projections of $\delta$ parallel and perpendicular to the super-weak direction, $\phi_{SW} = \tan^{-1}(2\Delta m/\Delta \gamma)$.

The projections $\delta_{\|}$ and $\delta_{\bot}$ are related to the mass and width difference between the strong interaction eigenstates $K^{0}$ and $\bar{K}^{0}$:
\beq
\label{eq:deltas}
\delta_{\|} = \frac{1}{4}\frac{ \gamma_{K^{0}}-\gamma_{\bar{K}^{0}}}{\sqrt{\Delta m^{2}+\Big(\frac{\Delta \gamma}{2}\Big)^{2}}},~~\delta_{\bot}=\frac{1}{2}\frac{m_{K^{0}}-m_{\bar{K}^{0}}}{\sqrt{\Delta m^{2}+\Big(\frac{\Delta \gamma}{2}\Big)^{2}}}.
\eeq

From Eqs. (\ref{eq:lowest-order1}) and (\ref{eq:lowest-order2}) we see that the super-weak angle $\phi_{SW}$ is not affected by noncommutativity to the lowest order. That is,
\bea
\phi_{SW} &=& \tan^{-1}(2\Delta m/\Delta \gamma)\nn \\
&\simeq& \tan^{-1}(\gamma_{K^{0}}/2m_{K^{0}}).
\eea

The real and imaginary parts of $\delta$ are known from kaon decay experiments. From \cite{Angelopoulos98F, LAI05A} we have 
\beq
\textrm{Re}~\delta \simeq 2.9 \times 10^{-4},~~\textrm{Im}~\delta \simeq -0.2 \times 10^{-5}.
\eeq  

From Eq. (\ref{eq:delta-eqn}), we have the expressions for $\delta_{\|}$ and $\delta_{\bot}$:
\bea
\delta_{\|} &=&\textrm{Re}~\delta~\cos(\phi_{SW}) + \textrm{Im}~\delta~\sin(\phi_{SW}),\\
\delta_{\bot}&=&-\textrm{Re}~\delta~\sin(\phi_{SW}) + \textrm{Im}~\delta~\cos(\phi_{SW}).
\eea

On using the super-weak angle measured at the KTeV E731 experiment \cite{Gibbons1997}
\beq
\phi_{SW} = 43.4^{\circ} \pm 0.1^{\circ},
\eeq
we obtain the values for $\delta_{\|}$ and $\delta_{\bot}$,
\beq
\delta_{\|} \simeq 20.93 \times 10^{-5},~~~\delta_{\bot}\simeq -20.07 \times 10^{-5}.
\eeq

From Eq. (\ref{eq:deltas}) we have the mass difference
\bea
m_{K^{0}}-m_{\bar{K}^{0}} &=& 2 \delta_{\bot}\sqrt{\Delta m^{2}+\Big(\frac{\Delta \gamma}{2}\Big)^{2}}\nn \\
&\simeq& \delta_{\bot}(m_{K^{0}}\vec{\theta}^{0}.\vec{P}_{in})\sqrt{\frac{1}{4}\gamma^{2}_{K^{0}}+m^{2}_{K^{0}}}\nn
\eea
On using $\textrm{tan}(\phi_{SW}) \simeq \gamma_{K^{0}}/2m_{K^{0}}$ we have
\beq
m_{K^{0}}-m_{\bar{K}^{0}} \simeq \delta_{\bot}(m_{K^{0}}\vec{\theta}^{0}.\vec{P}_{in}) m_{K^{0}}\sqrt{1 + \tan^{2}(\phi_{SW})}.
\eeq
Thus the CPT figure of merit takes the form 
\beq
r_{K^{0}} \equiv \frac{|m_{K^{0}}-m_{\bar{K}^{0}}|}{m_{K^{0}}} \simeq \delta_{\bot}(m_{K^{0}}\vec{\theta}^{0}.\vec{P}_{in}) \sqrt{1 + \tan^{2}(\phi_{SW})}.
\eeq

The noncommutativity parameter measured in the laboratory frame can in fact vary with time due to the rotation of the Earth. Let us denote the laboratory frame by ($\hat{x}, \hat{y}, \hat{z}$) and the non-rotating frame (compatible with the Earth's celestial equatorial coordinates) by ($\hat{X}, \hat{Y}, \hat{Z}$). The components of the noncommutativity parameter measured in the laboratory frame at time $t$, $(\theta^{x}(t), \theta^{y}(t), \theta^{z}(t))$ can be connected to the components in the non-rotating frame $(\theta^{X}, \theta^{Y}, \theta^{Z}) \equiv (\theta^{01}, \theta^{02}, \theta^{03})$. 

The relations connecting the components of a vector between these two frames are known in the literature. From \cite{Green} we have
\bea
\theta^{x}(t) &=& \theta^{X}\cos\chi\cos\Omega t+ \theta^{Y}\cos\chi \sin\Omega t-\theta^{Z}\sin\chi,~~~~~\\
\theta^{y}(t) &=& -\theta^{X}\sin\Omega t+ \theta^{Y}~\cos\Omega t,~~~~~~~\\
\theta^{z}(t) &=& \theta^{X}\sin\chi\cos \Omega t+ \theta^{Y}\sin\chi\sin\Omega t+\theta^{Z}\cos\chi,~~~~~
\eea
where $\chi = \cos^{-1}(\hat{z}.\hat{Z})$ and $\Omega$ is the sidereal frequency of the Earth.

In the laboratory frame we have the relation 
\beq
\vec{\theta}^{0}(t).\vec{P}_{in} \simeq \Big(\frac{|m_{K^{0}}-m_{\bar{K}}^{0}|}{m_{K^{0}}}\Big) \frac{1}{\delta_{\bot}m_{K^{0}}\sqrt{1 + \tan^{2}(\phi_{SW})}}.  
\eeq

The KTeV experiment at Fermilab involves highly collimated kaon beam with an average boost factor $\bar{\gamma}$ of the order of 100 and $\beta=v/c \simeq 1$. The $\hat{z}$-axis of the laboratory frame is chosen along the beam direction such that the kaon three-velocity reduces to $\vec{\beta} = (0, 0, \beta)$. In that case the above equation reduces to
\beq
\theta^{z}(t)P^{z}_{in} \simeq \Big(\frac{|m_{K^{0}}-m_{\bar{K}}^{0}|}{m_{K^{0}}}\Big) \frac{1}{\delta_{\bot}m_{K^{0}}\sqrt{1 + \tan^{2}(\phi_{SW})}}.
\eeq
In the non-rotating frame,
\bea
&&(\theta^{X}\cos \Omega t+ \theta^{Y}\sin\Omega t)\sin\chi +\theta^{Z}\cos\chi \nn \\
&&\simeq \Big(\frac{|m_{K^{0}}-m_{\bar{K}}^{0}|}{m_{K^{0}}}\Big) \frac{1}{\delta_{\bot}m_{K^{0}}P^{z}_{in}\sqrt{1 + \tan^{2}(\phi_{SW})}}.~~~ 
\eea

The first two terms in the left hand side oscillate in time with a frequency $\Omega$. Since experiments are performed over extended time periods, we may disregard the time dependence, and thus take the time averaged form of the above expression. Thus we have
\beq
\theta^{Z}\cos\chi \simeq \Big(\frac{|m_{K^{0}}-m_{\bar{K}}^{0}|}{m_{K^{0}}}\Big) \frac{1}{\delta_{\bot}(m_{0K^{0}})^2\bar{\gamma}^2\bar{\beta}c\sqrt{1 + \tan^{2}(\phi_{SW})}}, 
\eeq
where we have replaced the incident momentum $P^{z}_{in}$ by $m_{0K^{0}}\bar{\gamma}\bar{\beta}c$, $m_{K^{0}}$ by $m_{0K^{0}}\bar{\gamma}$, with $m_{0K^{0}}$ the kaon rest mass. $\bar{\gamma}$ and $\bar{\beta}$ are averages of $\beta$ and $\gamma$. This expression gives a bound for the $z$-component of the noncommutativity parameter in the non-rotating frame.

The detector geometry in the KTeV experiment has $\textrm{cos}\chi=0.6$. Thus on using the results from the experiments on kaons we obtain
\bea
\theta^{Z} &\simeq& \Big(\frac{|m_{K^{0}}-m_{\bar{K}}^{0}|}{m_{K^{0}}}\Big) \frac{\hbar^{2}/c}{\delta_{\bot}\textrm{cos}\chi~(m_{0K^{0}})^2\bar{\gamma}^2\bar{\beta}c\sqrt{1 + \tan^{2}(\phi_{SW})}}\nn \\
&\lesssim&0.0829 \times 10^{-63} \textrm{m}^{2}.
\eea
This gives an upper bound for the noncommutativity length scale 
\beq
\sqrt{\theta^{Z}} \lesssim 10^{-32} \textrm{m},
\eeq
corresponding to a lower bound for energy $E$ associated with spacetime noncommutativity
\beq
E \gtrsim  10^{16}\textrm{GeV}.
\eeq
\section{Noncommutativity bound from $g-2$ difference of $\mu^{+}$ and $\mu^{-}$}\label{g-minus-two}
We can put further constraint on the noncommutativity length scale from the CPT violation measurements on the $g-2$ of positive and negative muons \cite{:2007yc, Bailey, Carey, Brown}. (It is also possible to put a bound on spacetime noncommutativity from the $g-2$ difference between electron and positron \cite{Vandyck, Dehmelt}. It turns out that this bound does not constrain the noncommutativity parameter very well due to the small mass of the electron.)

\begin{figure}
\begin{center}
\includegraphics{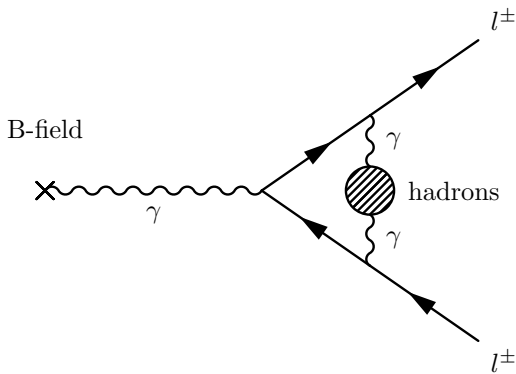}
\caption{Lowest order hadronic contribution to $g-2$ of a lepton $l^{\pm}$ ($\mu^{\pm}$ or $e^{\pm}$) in a uniform magnetic field $B$.} \label{fig:2}
\end{center}
\end{figure}

In twisted noncommutative field theories on the Groenewold-Moyal (GM) plane (noncommutative spacetime modeled by the star-product approach with deformed statistics) gauge-matter field vortices are in general affected by spacetime noncommutativity. It is shown elsewhere \cite{Balachandran:2006pi} that the $S$-operator in an abelian gauge-matter theory (say, QED) is unaffected by spacetime noncommutativity. However, higher order hadronic (and thus non-abelian) loop corrections to the QED scattering diagrams can in fact carry a non-trivial dependence on spacetime noncommutativity. Such a diagram is pictured in Fig. \ref{fig:2}, where the $g-2$ of a lepton $l^{\pm}$ ($\mu^{\pm}$ or $e^{\pm}$) receives a lowest order hadronic loop contribution when it is moving in an external electromagnetic field. The experiments performed at CERN and BNL \cite{:2007yc, Bailey, Carey, Brown} to measure the muon $g-2$ use a storage ring magnet in which muons are circulating in a uniform magnetic field. (The Penning trap experiments to determine electron-positron $g$-factors \cite{Vandyck, Dehmelt} use a strong homogeneous magnetic field and a quadrupole electric field.)

The interaction Hamiltonian density of the noncommutative field theory in general splits into two parts, a part with matter-gauge couplings and pure matter coupling ${\cal H}_{I(\theta)}^{M, G}$ and a pure gauge part ${\cal H}_{I(\theta)}^{G}$ \cite{Akofor:2007hk, Balachandran:2007yf},
\beq
{\cal H}_{I(\theta)} = {\cal H}_{I(\theta)}^{M, G} + {\cal H}_{I(\theta)}^{G},
\eeq
where 
\beq
{\cal H}_{I(\theta)}^{M, G} = {\cal H}_{I(0)}^{M, G}e^{\frac{1}{2}\overleftarrow{\partial} \wedge P},~~~{\cal H}_{I(\theta)}^{G}={\cal H}_{I(0)}^{G}
\eeq

In a nonabelian gauge theory, ${\cal H}^{G}_{I(\theta)} = {\cal H}^{G}_{I(0)}\neq 0$, so that the $S$-operator $S^{M,G}_{(\theta)}$ of the theory
\beq
S^{M,G}_{(\theta)} \neq S_{(\theta =0)}^{M,G} = S^{M,G}.
\eeq
This can in fact affect the lepton-photon vertex function through the contribution from hadronic (and thus non-abelian) loops. (See Fig. \ref{fig:2}.) 

The $S$-operator $S^{M,G}_{(\theta)}$ depends only on $\theta^{0i}$ \cite{Akofor:2007hk},
\beq
S^{M,G}_{(\theta)} = S_{\theta^{0i}}^{M,G}.
\eeq

Charge conjugation C and time reversal T on $S^{MG}_{(\theta)}$ do not affect $\theta^{0i}$, while parity P changes its sign \cite{Akofor:2007hk}. Thus a nonzero $\theta^{0i}$ contributes to P and thus CPT violation.

The appearance of the term $\theta^{0i}P_{i}^{\textrm{inc}}$, where $P_{i}^{\textrm{inc}}$ is the total incident momentum, in the $S$-operator suggests that the amount of CPT violation to the leading order in $\theta^{\mu \nu}$ should be ${\cal O}(\vec{\theta}^{0}\cdot {\overrightarrow P}_{\textrm{inc}})$.

From the CERN experiments on positive and negative muon $g$-factors, the standard CPT figure of merit is given by \cite{Bailey}
\beq
\label{eq:for-muon}
r_{g}^{\mu} \equiv \frac{|g_{{\mu}^{+}} - g_{{\mu}^{-}}|}{g^{avg}_{\mu}} \lesssim 10^{-8}.
\eeq

To the leading order in $\theta^{\mu \nu}$
\beq
r_{g}^{\mu} \approx m^{avg}_{\mu}\vec{\theta}^{0}\cdot {\overrightarrow P}_{\textrm{inc}},
\eeq
where $m^{avg}_{\mu}$ is the average mass of the positive- and negative muons, the only relevant mass scale in the theory.

We get the maximum bound when $\vec{\theta}^{0}$ and ${\overrightarrow P}_{\textrm{inc}}$ are parallel. In that case
\beq
\label{eq:muon-bound}
|\vec{\theta}|\lesssim \frac{10^{-8}}{(m_{\mu}\gamma)^{2}}~,
\eeq
where $\gamma$ is the relativistic factor. The muon $g-2$ experiments at CERN and BNL were performed at a specific value of the relativistic factor, $\gamma = 29.3$.

Eq. (\ref{eq:muon-bound}) gives an upper bound for the length scale of noncommutativity
\bea
|\vec{\theta}| &\lesssim& \frac{10^{-8}}{(m_{\mu}\gamma)^{2}}\simeq1.61 \times 10^{-39}m^{2} \rightarrow \sqrt{\theta} \lesssim 10^{-20} m.~~~~
\eea
This corresponds to a lower bound for the energy scale $E \gtrsim 10^3$ GeV. 

The Penning trap experiments measure the difference in electron and positron $g$-factors \cite{Dehmelt, Vandyck}, $r_{g}^{e} \equiv |g_{{e}^{+}} - g_{{e}^{-}}|/g^{avg}_{e} \lesssim 10^{-12}$. This CPT figure of merit is more precise compared to that of the muon. However it is much less sensitive to the new physics arising from spacetime noncommutativity as the electron has lower mass compared to muons and kaons. 

\section{Noncommutative corrections to elastic electron-nucleon scattering}\label{form-factors}
In this section we study how the spacetime noncommutativity affects the elastic electron-nucleon scattering process at the tree level. We show that the nucleon vertex function and the electromagnetic form factors are affected by noncommutativity. The spatial distributions of the charge and magnetization carried by the nucleon are also modified.
\subsection{Phenomenology of Electron-Nucleon Interaction on the Commutative Spacetime}
\begin{figure}
\begin{center}
\includegraphics[height=4.5cm]{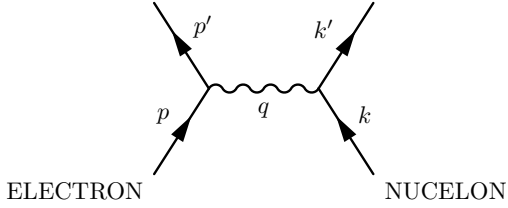}
\caption{Feynman diagram of electron-nucleon scattering caused by the exchange of a virtual photon.} \label{fig:1}
\end{center}
\end{figure}

The scattering process of an electron from a nucleon through the exchange of a virtual photon is represented by the Feynman diagram in Fig. \ref{fig:1}. Such a process is contained in the matrix element
\bea
\label{eq:matrix}
\langle p', k'|S^{(2)}|p, k \rangle &=& \langle p', k'|{\bf T}\Big(~\frac{(-i)^2}{2!}\int d^4 x d^4 y~j^{p,n}_{\mu}(x)\nn \\
&&\times A^{\mu}(x)j^{e}_{\mu}(y)A^{\mu}(y)~\Big)|p, k \rangle,
\eea
where $j^{p,n}_{\mu}$, $j^{e}_{\mu}$, respectively are the electron- and nucleon currents, $A^{\mu}$ is the photon field and $p$, $p'$ and  $k$, $k'$ are the four-momenta of the incident and scattered electron and nucleon respectively.

In momentum space, up to a possible minus sign, it takes the form
\bea
\label{eq:matrix-element}
&&\int \frac{d^4 q}{(2\pi)^4}~(2\pi)^4 \delta^{(4)}(p'-p+q)~(2\pi)^4 \delta^{(4)}(k'-k-q)\nn \\
&&\times j^{p,n}_{\mu}(p',p)~\frac{ig^{\mu \nu}}{q^2}~j^{e}_{\nu}(k',k),
\eea
where $g^{\mu \nu}$ is the metric tensor. The factor $1/q^2$ represents the propagation of a virtual photon of four-momentum $q_{\mu}$ between the electron and nucleon. It is
\beq
q_{\mu} = (p'_{\mu} - p_{\mu}) =-(k'_{\mu} - k_{\mu}).
\eeq

This expression takes the form $i {\cal M} (2\pi)^4 \delta^{(4)}(p'+k'-p-k)$, after a delta function integration, with
\beq
i {\cal M} = j^{p,n}_{\mu}(p',p)\Big(\frac{ig^{\mu \nu}}{q^2}\Big)j^{e}_{\nu}(k',k).
\eeq

The electron charge-current density $j_{\mu}^{e}$, which, assuming that the electron has no internal structure, is given by
\beq
j_{\mu}^{e}(p', p) = -ie~\bar{u}(p') \gamma_{\mu} u(p),
\eeq
where $\bar{u}$ and $u$ are the Dirac spinors of the electron. 

The nucleon charge-current density $j_{\mu}^{p, n}$ (proton or neutron) is given by
\beq
j_{\mu}^{p, n}(k', k) =-ie~\bar{N}(k')\Gamma^{\mu}(k', k)N(k),
\eeq
where $\bar{N}$ and $N$ are the Dirac spinors of the nucleon and $\Gamma^{\mu}$, called the vertex function, includes all the effects due to the internal structure of the nucleon.

In terms of the electron- and nucleon spinors, the matrix element for elastic scattering takes the form
\beq
-i {\cal M} =\frac{-ig_{\mu \nu}}{q^2}~\Big(ie \bar{u}(p')\gamma^{\nu}u(p)\Big)\Big(-ie \bar{N}(k')\Gamma^{\mu}(k', k)N(k)\Big).
\eeq

If we assume that (i) $j_{\mu}^{p, n}$ transforms as a four-vector (relativistic covariance), (ii) $j_{\mu}^{p, n}$ is conserved and (iii) the nucleon is a Dirac particle, then the nucleon charge-current density is constrained to be of the form \cite{Foldy, Salzman} 
\bea
\label{eq:nucleon-charge-current}
j_{\mu}^{p, n} (p', p) &=& ie \bar{N}(p') \Big[\gamma_{\mu} F_{1}^{p, n}(q^2) + \frac{\kappa^{p, n}}{2m^{p, n}} \nn \\
&&\times \sigma_{\mu \nu} q_{\nu} F_{2}^{p, n}(q^2)\Big]N(p).
\eea
 
In Eq. (\ref{eq:nucleon-charge-current}), $\kappa^{p, n}$ is the anomalous magnetic moment of the nucleon in nuclear magnetons, $m^{p, n}$ is the mass of the nucleon, $\sigma_{\mu \nu} = (1/2i) [\gamma_{\mu}, \gamma_{\nu}]$, and the functions $F_{1, 2}^{p, n}$ (known as the electromagnetic form factors) describe the internal structure of the nucleon.

The form factors essentially measure how strongly the nucleon `holds together' and recoils when a momentum $q_{\mu}$ is exchanged in the scattering process. They give the spatial distributions of charge and magnetic moment of the nucleon, $G_{E_{p, n}}$ and $G_{M_{p, n}}$ respectively,
\bea
G_{E_{p, n}} (q^2) &=& F_{1}^{p, n}(q^2) - \frac{q^2}{4m_{p, n}^2} F_{2}^{p, n}(q^2)\\
G_{M_{p, n}} (q^2) &=& F_{1}^{p, n}(q^2) + F_{2}^{p, n}(q^2)
\eea
$G_{E_{p}}$, $G_{M_{p}}$, $G_{E_{n}}$, $G_{M_{n}}$ are called the electric- and magnetic Sachs form factors for the proton and neutron respectively \cite{Perdrisat, Ernst, Sachs}.  
\subsection{Phenomenology of Electron-Nucleon Interaction on the Noncommutative Spacetime}
Matter fields on the GM plane obey twisted (statistics deformed) commutation relations \cite{Balachandran:2007kv, Balachandran:2007vx, Balachandran:2006pi}. The noncommutative spinor field $\psi^{(\theta)}$ is related to its commutative counterpart $\psi$ through the relation \cite{Balachandran:2007kv, Akofor:2007hk}
\beq
\label{eq:spinor}
\psi^{(\theta)}({\bf x}, t) = \psi({\bf x}, t) e^{\frac{1}{2}\overleftarrow{\partial} \wedge P},
\eeq
where $\overleftarrow{\partial} \wedge P = \overleftarrow{\partial}_{\mu} \theta^{\mu \nu} P_{\nu}$ and $P_{\nu}$ is the total momentum operator. 

Matter fields on the GM plane must be transported by the connection compatibly with Eq. (\ref{eq:spinor}). It imposes a natural choice on the covariant derivatives \cite{Balachandran:2007kv, Akofor:2007hk}, making the gauge sector of the theory commutative. Thus the gauge field $A_{\mu}^{(\theta)}({\bf x}, t)$ on the GM plane is the same as its commutative counterpart
\beq
A^{(\theta)}_{\mu}({\bf x}, t) = A_{\mu}({\bf x}, t).
\eeq

The Feynman rules for arbitrary noncommutative scattering processes are investigated in \cite{Bal2009}. We focus, here, on a simple way to find the noncommutative corrections to the tree level electron-nucleon elastic scattering process. 

Since we are considering a $U(1)$ gauge theory, we have ${\cal H}_{I(\theta)}^{G}={\cal H}_{I(0)}^{G}=0$. It is shown elsewhere \cite{Balachandran:2006pi, Balachandran:2005pn, Balachandran:2007yf} that the $S$-operators of such theories are the same as those of their commutative counterparts. In an abelian gauge theory such as QED,
\beq
S_{(\theta)} = S_{(\theta = 0)} = S.
\eeq

To find the $S$-matrix element for the electron-nucleon scattering at the tree-level, we look at the noncommutative version of the matrix element given in Eq. (\ref{eq:matrix}),
\bea
\label{eq:matrix-theta}
{\ }_{(\theta)}\langle p', k'|S_{(\theta)}^{(2)}|p, k \rangle_{(\theta)} &=& {\ }_{(\theta)}\langle p', k'|S^{(2)}|p, k \rangle_{(\theta)}
\eea

The noncommutative matrix element differs from its commutative counterpart through the appearance of twisted incident- and outgoing state vectors.

We write down the twisted incident- and outgoing state vectors in terms of the twisted fields acting on the vacuum
\beq
|p,~k\rangle_{(\theta)}=a^{\dagger}_{N}(p)a^{\dagger}_{e}(k)|0\rangle,~{\ }_{(\theta)}\langle p',~k'| = \langle 0|a_{N}(p')a_{e}(k'). 
\eeq

The map from noncommutative creation- and annihilation operators $a^{\dagger}_{e, N}(k)$, $a_{e, N}(k)$ to the corresponding commutative creation- and annihilation operators $c^{\dagger}_{e, N}(k)$, $c_{e, N}(k)$ (called the ``dressing transformation") can be used to untwist the incident- and outgoing state vectors. The map is \cite{Balachandran:2007vx}
\beq
a^{\dagger}_{e, N}(k) = c^{\dagger}_{e, N}(k) e^{\frac{i}{2}k \wedge P},~a_{e, N}(k) = c_{e, N}(k) e^{-\frac{i}{2}k \wedge P}.
\eeq 

The twisted incident- and outgoing state vectors can be expressed in terms of the untwisted incident- and outgoing state vectors
\bea
\label{eq:untwist1}
|p,~k\rangle_{(\theta)}&=&e^{\frac{i}{2}p \wedge k}c^{\dagger}_{N}(p)c^{\dagger}_{e}(k)|0\rangle=e^{\frac{i}{2}p \wedge k}|p,~k\rangle
\eea
and
\bea
\label{eq:untwist2}
{\ }_{(\theta)}\langle p',~k'| &=& e^{-\frac{i}{2}p' \wedge k'}\langle 0|c_{N}(p')c_{e}(k')\nn \\
&=&e^{-\frac{i}{2}p' \wedge k'}\langle p',~k'|
\eea

The second order term in the $S$-operator expansion $S^{(2)}_{(\theta)}$ is independent of $\theta$ \cite{Balachandran:2006pi, Balachandran:2005pn, Balachandran:2007yf}. Hence 
\beq
S^{(2)}_{(\theta)} = S^{(2)}_{(\theta=0)}=S^{(2)},
\eeq
the commutative $S$-operator.

We write down $S$-matrix element in noncommutative spacetime (given in Eq. (\ref{eq:matrix-theta})) in terms of the commutative matrix element,
\beq
\label{eq:noncommu-matrix2}
{\ }_{(\theta)}\langle p', k'|S_{(\theta)}^{(2)}|p, k \rangle_{(\theta)}=e^{-\frac{i}{2}p' \wedge k'}e^{\frac{i}{2}p \wedge k}\langle p', k'|S^{(2)}|p, k \rangle.
\eeq

Thus the noncommutative matrix element is
\bea
i {\cal M}_{(\theta)} &=& e^{-\frac{i}{2}p' \wedge k'}e^{\frac{i}{2}p \wedge k}~j^{p,n}_{\mu}(p',p)~\Big(\frac{ig^{\mu \nu}}{q^2}\Big)~j^{e}_{\nu}(k',k)\nn \\
&=& \frac{-ig_{\mu \nu}e^{\frac{i}{2}(p+k)\wedge q}}{q^2}\Big[ie \bar{u}(k')\gamma^{\nu}u(k)\Big]\nn \\
&&\times \Big[-ie \bar{N}(p')\Gamma^{\mu}(p', p)N(p)\Big]~~
\eea

In the tree level scattering process we assumed that the electron is a point particle and the nucleon has an internal structure. The vertex function $\Gamma^{\mu}$ contains all the details of the internal structure of the nucleon. We infer that the additional $\theta^{\mu \nu}$ dependent factor represents the noncommutative modification of the internal structure of the nucleon.

The nucleon charge-current density given in Eq. (\ref{eq:nucleon-charge-current}) is effectively modified in the noncommutative case, and it takes the form
\bea
\label{eq:nucleon-charge-current1}
j_{\mu}^{p, n (\theta)} (k, p, q) &=& ie~e^{\frac{i}{2}(p+k)\wedge q} \bar{N}(p') \Big[\gamma_{\mu} F_{1}^{p, n}(q^2) + \frac{\kappa^{p, n}}{2m}\nn \\
&& \times \sigma_{\mu \nu} q_{\nu} F_{2}^{p, n}(q^2)\Big]N(p).
\eea

This shows that the electromagnetic form factors are modified
\beq
F_{1,2}^{p, n (\theta)}(k, p, q) = e^{\frac{i}{2}(p+k)\wedge q} F_{1,2}^{p, n}(q^2). 
\eeq
They are dependent on the total incident four-momentum $p_{\mu}+k_{\mu}$ and the recoil four-momentum $q_{\nu}$ of the scattering particles and the noncommutativity parameter $\theta^{\mu \nu}$. 

In the noncommutative case, the spatial distributions of charge and magnetic moment of the nucleon (Sachs form factors), $G^{(\theta)}_{E_{p, n}}$ and $G^{(\theta)}_{M_{p, n}}$ respectively are,
\bea
G^{(\theta)}_{E_{p, n}} (k, p, q) &=& e^{\frac{i}{2}(p+k)\wedge q}\Big(F_{1}^{p, n}(q^2)- \frac{q^2}{4m_{p, n}^2} F_{2}^{p, n}(q^2)\Big)~~~~~~\\
G^{(\theta)}_{M_{p, n}} (k, p, q) &=& e^{\frac{i}{2}(p+k)\wedge q}\Big(F_{1}^{p, n}(q^2) + F_{2}^{p, n}(q^2)\Big).~~~~~
\eea
They are now functions of $k$, $p$, $q$ and direction-dependent, unlike the commutative case. Possible experimental signals due to these effects should be explored further.
\section{Conclusions}\label{conclusions}
In this paper, we have constrained the spacetime noncommutativity parameter using the CPT figure of merit measured in the $K^{0} - \bar{K}^{0}$ system and $g-2$ difference of positive- and negative muons following a phenomenological approach. We get the noncommutativity length scale bounds $\lesssim10^{-32}$m and $\lesssim10^{-20}$m and energy bounds $\gtrsim10^{16}$GeV and $\gtrsim10^{3}$GeV respectively from $K^{0} - \bar{K}^{0}$ system and $g-2$ of $\mu^{+}-\mu^{-}$.  

We have also shown that the electromagnetic form factors and thus the distributions of charge and magnetization of the nucleon are modified. The form factors are no longer analytic functions of $q^2$ as in the commutative case. They are direction dependent in the noncommutative case, indicating the possibility of Lorentz violation. It may lead to interesting experimental signals.
\section{Acknowledgements}
I wish to thank A. P. Balachandran for suggesting the problem, numerous helpful discussions and carefully reading the manuscript. I also wish to thank Earnest Akofor, Mario Martone, Pramod Padmanabhan and Paul Souder for useful discussions and Don Bunk for carefully reading the earlier version of the manuscript. This work was supported in part by the US Department of Energy grant under the contract number DE-FG02-85ER40231.

\end{document}